\documentclass [prl,twocolumn,superscriptaddress]{revtex4-1}

\usepackage{graphicx}
\usepackage{subfigure}
\usepackage{amsmath}
\usepackage{mathrsfs}
\usepackage{amsthm}
\usepackage{amssymb}
\usepackage{hyperref}
\usepackage{xcolor}
\usepackage{csquotes}

\MakeOuterQuote{"}
\hypersetup{colorlinks=true, linkcolor=blue, citecolor=blue, urlcolor=blue}
\allowdisplaybreaks

\newcommand{\ket}[1]{\left| #1 \right>} 
\newcommand{\bra}[1]{\left< #1 \right|} 

\begin{document}

\title{Disorder-Driven Density and Spin Self-Ordering of a Bose-Einstein Condensate in a Cavity}
 
\author{Farokh Mivehvar}
\email[Corresponding author: ]{farokh.mivehvar@uibk.ac.at}
\affiliation{Institut f\"ur Theoretische Physik, Universit{\"a}t Innsbruck, A-6020~Innsbruck, Austria}
\author{Francesco Piazza}
\affiliation{Max-Planck-Institut f\"{u}r Physik komplexer Systeme, D-01187 Dresden, Germany}
\author{Helmut Ritsch}
\affiliation{Institut f\"ur Theoretische Physik, Universit{\"a}t Innsbruck, A-6020~Innsbruck, Austria}

\begin{abstract}
We study spatial spin and density self-ordering of a two-component
Bose-Einstein condensate via collective Raman scattering into a linear
cavity mode. The onset of the Dicke superradiance phase transition is
marked by a simultaneous appearance of a crystalline density order and
a spin-wave order. The latter spontaneously breaks the discrete
$\mathbf{Z}_2$ symmetry between even and odd sites of the cavity
optical potential. Moreover, in the superradiant state the continuous
$U(1)$ symmetry of the relative phase of the two condensate
wavefunctions is explicitly broken by the cavity-induced
position-dependent Raman coupling with a zero spatial average. Thus,
the spatially-averaged relative condensate phase is locked at either $\pi/2$ or $-\pi/2$.
This continuous symmetry breaking and relative condensate phase locking by a zero-average Raman field can be considered as a generic order-by-disorder process similar to the random-field-induced order in the two-dimensional classical ferromagnetic $XY$ spin model. However, the seed of the random field in our model stems from quantum fluctuations in the cavity field and is a dynamical entity affected by self-ordering. The spectra of elementary excitations exhibit the typical mode softening at the superradiance threshold.
\end{abstract}

\maketitle

\emph{Introduction.}---Loading Bose-Einstein condensates (BECs) 
into optical potentials created by dynamic cavity fields has opened a new avenue in ultracold atomic 
physics~\cite{Brennecke2007,Slama2007,Colombe2007,Kollar2015}, paving the way for realization of 
novel phenomena~\cite{Ritsch2013}. 
Seminal results include the Dicke superradiance phase 
transition~\cite{Baumann2010,Baumann2011,Klinder2015}, 
and quantum phase transitions between superfluid, 
superradiant Mott insulator, density-wave state, lattice supersolid, and supersolid phase with a broken 
continuous $U(1)$ symmetry due to the interplay between cavity-mediated long-range interactions and 
short-range collisional interactions~\cite{Klinder2015b,Landig2016,Leonard2017}.  
On the theoretical side, in addition to studying conventional quantum-optics and self-ordering aspects of coupled 
quantum-gas--cavity environments~\cite{Moore1999,Domokos2002,Maschler2008,Nagy2008,
Piazza2013,Piazza2014,Keeling2014,Chen2014,Piazza2015},  many proposals have been put forward 
to simulate and realize exotic phenomena for ultracold atoms via coupling to dynamic cavity fields, 
including synthetic gauge fields~\cite{Mivehvar2014, Dong2014,Deng2014,Mivehvar2015,Kollath2016,Zheng2016}, 
topological states~\cite{Pan2015,Gulacsi2015,Mivehvar2017,Sheikhan2016b},
and superconductor-related physics~\cite{Ballantine2017}.

\begin{figure}[b!]
\centering
\includegraphics [width=0.46\textwidth]{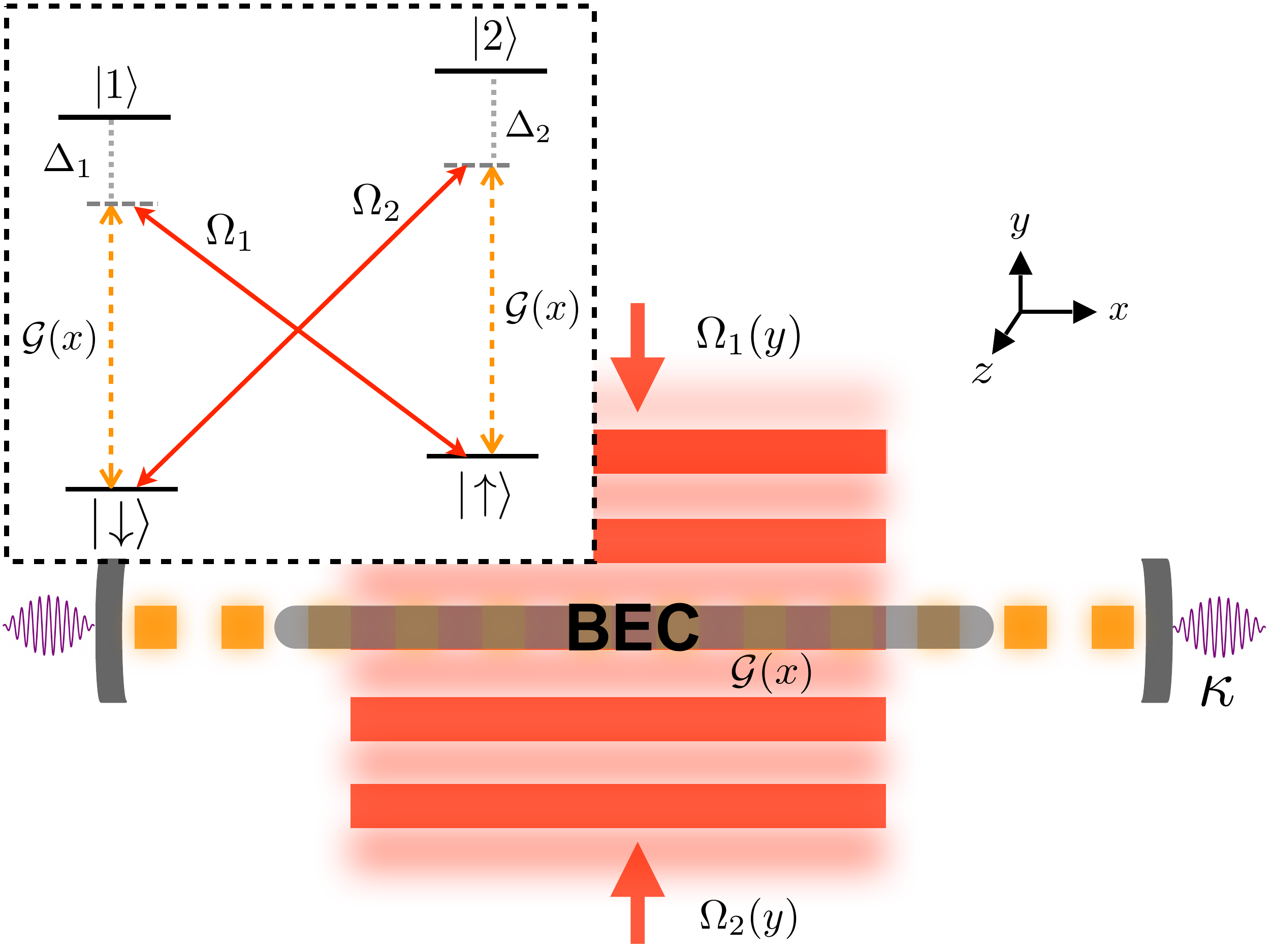}
\caption{Schematic view of a transversely pumped one-dimensional spinor BEC inside a cavity. 
The inset depicts the internal atom-photon coupling scheme in double $\Lambda$ configuration. 
The first (second) pump laser solely induces the transition $\uparrow\>\leftrightarrow1$ ($\downarrow\>\leftrightarrow2$)
with the Rabi frequency $\Omega_1$ ($\Omega_2$), while the transition $\downarrow\>\leftrightarrow 1$ 
and $\uparrow\>\leftrightarrow 2$ are coupled to the cavity mode with the identical strength $\mathscr{G}(x)$.} 
\label{fig:lin-cavity-geom--Lambda-scheme}
\end{figure}

In this Letter we study the Dicke superradiance phase transition for a generalized atomic system  
with both internal~\cite{Dimer2007,Larson2009,Gopalakrishnan2011,Safaei2013,Zhiqiang2017} and 
external~\cite{Baumann2010,Baumann2011,Klinder2015} quantized degrees of freedom, 
i.e., a spinor BEC, coupled to a single mode of a linear cavity 
(see Fig.~\ref{fig:lin-cavity-geom--Lambda-scheme}).
The ultracold four-level atoms are transversely illuminated by two sufficiently far red-detuned 
pump lasers polarized along the cavity axis $x$ so that to induce near resonant two-photon Raman 
transitions between the lowest two internal atomic states via the same cavity mode 
with the transverse polarization along $z$ as in Ref.~\cite{Zhiqiang2017}. 
After adiabatic elimination of the atomic excited states, the system reduces
to a two-component BEC coupled via a cavity-induced position-dependent Raman coupling 
with a zero spatial average. 

In contrast to conventional self-ordering~\cite{Ritsch2013,Baumann2010}, 
the condensate densities in the superradiant phase exhibit modulations with the
half cavity-wavelength $\lambda_c/2$ periodicity. However, the discrete $\mathbf{Z}_2$ symmetry --- the 
symmetry between even and odd lattice sites and positive and negative cavity-field 
amplitude --- is \textit{spontaneously} broken at the onset of the Dicke superradiance 
phase transition by a $\lambda_c$-periodic spin ordering. 
Despite filling all sites, the cavity field attains a non-zero value as 
it is collectively driven by the atomic spin density. 
The continuous $U(1)$ symmetry associated with the freedom of
the relative phase of two condensate wavefunctions is \textit{explicitly} broken in the superradiance state
by the cavity-induced position-dependent Raman coupling with the zero spatial average.
Thereby the relative condensate phase varies in space with the spatial average of 
either $\pi/2$ or $-\pi/2$ in order to minimize the total energy and yield a non-zero cavity field~\cite{Jacksh_2001}. 

Owing to this continuous symmetry breaking and relative condensate phase 
locking by the zero-average Raman field, the self-organization in our model can be considered
as an order-by-disorder process~\cite{Niederberger2008}, equivalent to 
the spontaneous ordering in the two-dimensional classical ferromagnetic $XY$ spin model 
with a uniaxial random magnetic field~\cite{Minchau1985,Feldman1998,Wehr2006}.
The relative condensate phase plays the role of the spin angle  and the 
position-dependent Raman field with zero spatial average mimics the 
random magnetic field.  Nonetheless, the seed of the random field in 
our model stems from quantum fluctuations and the random field itself is a dynamical entity affected by the self-ordering.


\emph{Model.}---Consider four-level bosonic atoms inside a linear cavity
illuminated in the transverse direction by two external standing-wave pump lasers 
as depicted in Fig.~\ref{fig:lin-cavity-geom--Lambda-scheme}. A tight confinement 
along the transverse directions is assumed to freeze transverse motion. 
Two-photon Raman coupling is induced via the transition 
$\uparrow\>\leftrightarrow1$ ($\downarrow\>\leftrightarrow2$) coupled to the first (second) external pump 
laser with the Rabi frequency $\Omega_1$ ($\Omega_2$), along with the transitions 
$\downarrow\>\leftrightarrow 1$ and $\uparrow\>\leftrightarrow 2$ coupled to same empty 
cavity mode with identical coupling strength  $\mathscr{G}(x)=\mathscr{G}_0\cos(k_cx)$.
This constitutes a double $\Lambda$ configuration, where $\ket{\tau}=\{\ket{\downarrow},\ket{\uparrow}\}$ 
are the desired ground pseudospin states and $\{\ket{1},\ket{2}\}$ are electronic excited states, with energies 
$\{\hbar\omega_\downarrow=0, \hbar\omega_\uparrow,\hbar\omega_{1},\hbar\omega_{2}\}$. 
The pump and cavity frequencies, respectively, $\{\omega_{p1},\omega_{p2}\}$ and $\omega_c$ 
are assumed to be far red detuned from the atomic transition frequencies,
that is, $\Delta_1\equiv(\omega_{p1}+\omega_{p2})/2-\omega_1$ and
$\Delta_2\equiv\omega_{p2}-\omega_2$ are large.
However, two-photon Raman transitions are close to resonant:
$\omega_c-\omega_{p1}\approx\omega_{2p}-\omega_c\approx\omega_\uparrow$.
The excited states then quickly reach steady states with negligible populations and can 
be adiabatically eliminated to obtain an effective model describing the two atomic pseudospin states
coupled to the cavity field.

At the mean-field level, the system is described by a set of three coupled
equations for the cavity-field amplitude $\alpha(t)=|\alpha(t)|e^{i\phi_\alpha(t)}$ and atomic condensate wavefunctions 
$\psi_{\tau}(x,t)=\sqrt{n_\tau(x,t)}e^{i\phi_\tau(x,t)}$~\cite{SM-Mivehvar2017}: 
\begin{align} \label{eq:coupled-eqs}
i\frac{\partial}{\partial t}\alpha&=\Big[-\Delta_c-i\kappa
+\sum_{\tau=\downarrow,\uparrow}U_\tau\int \cos^2(k_cx)n_\tau dx\Big]\alpha\nonumber\\
&+\eta\int\cos(k_cx)
\left(\psi_\downarrow^*\psi_\uparrow+\psi_\uparrow^*\psi_\downarrow\right) dx,\nonumber\\
i\frac{\partial}{\partial t}\psi_\downarrow&=
\Big[-\frac{\hbar}{2m}\frac{\partial^2}{\partial x^2}
+U_\downarrow|\alpha|^2\cos^2(k_cx)\Big]\psi_\downarrow\nonumber\\
&+\eta(\alpha+\alpha^*)\cos(k_cx)\psi_\uparrow,\nonumber\\
i\frac{\partial}{\partial t}\psi_\uparrow&=
\Big[-\frac{\hbar}{2m}\frac{\partial^2}{\partial x^2}+\tilde{\delta}+U_\uparrow|\alpha|^2\cos^2(k_cx)\Big]\psi_\uparrow\nonumber\\
&+\eta(\alpha+\alpha^*)\cos(k_cx)\psi_\downarrow.
\end{align}
Here, $\tilde{\delta}\equiv\omega_\uparrow-(\omega_{p2}-\omega_{p1})/2+\Omega_1^2/\Delta_1-\Omega_2^2/\Delta_2$ 
is the Stark-shifted two-photon detuning, $U_{\downarrow(\uparrow)}\equiv\mathscr{G}_0^2/\Delta_{1(2)}$ is the maximum depth
of the cavity-generated optical potential per photon for the spin-down (spin-up) atoms 
[or the maximum cavity-frequency shift per a spin-down (spin-up) atom], 
$\eta\equiv\mathscr{G}_0\Omega_1/\Delta_1=\mathscr{G}_0\Omega_2/\Delta_2$ is
the balanced Raman-Rabi frequency (or the effective cavity pump strength),
and $\Delta_c\equiv(\omega_{p1}+\omega_{p2})/2-\omega_c$.
The decay of the cavity mode has been modeled 
as the damping term $-i\hbar\kappa\alpha$, with $\kappa$ being the decay rate.
Two-body contact interactions between atoms 
have been assumed to be negligible for the sake of simplicity.
When $\alpha\neq0$, 
only the total number of the particles $N=\sum_\tau\int n_\tau(x) dx=\sum_\tau N_\tau$ is conserved
due to the effective Raman coupling between the two condensates.

The total energy of the system can be expressed as 
$E=-\hbar\Delta_c|\alpha|^2+\int\mathscr{E}(x)dx$, 
where the energy-functional density is given by 
\begin{align} \label{E-functional-density}
\mathscr{E}(x)&=
\frac{\hbar^2}{2m}\left(|\partial_x\psi_\downarrow|^2+|\partial_x\psi_\uparrow|^2\right)
+\hbar\tilde{\delta}n_\uparrow\nonumber\\
&+\hbar |\alpha|^2\cos^2(k_cx) (U_\downarrow n_\downarrow+U_\uparrow n_\uparrow)\nonumber\\
&+4\hbar\eta|\alpha|\sqrt{n_\downarrow n_\uparrow}\cos(k_cx)\cos{\phi_\alpha}\cos{\Delta\phi},
\end{align}
with $\Delta\phi\equiv\phi_\downarrow-\phi_\uparrow$. As can be seen from Eq.~\eqref{E-functional-density},
in the absence of $\alpha$ the system possesses full translation symmetry $\mathcal{T}$
and $U(1)\times U(1)$ global symmetry representing the freedom of total and relative phases
of the two condensate wavefunctions. A non-zero cavity field results in a position-dependent Raman field 
with zero mean $\int_{0}^{\lambda_c} \eta(\alpha+\alpha^*)\cos{(k_cx)} dx=0$ over one 
cavity wavelength $\lambda_c=2\pi/k_c$, and reduces
the $\mathcal{T}\times U(1)\times U(1)$ symmetry into a $\mathbf{Z}_2\times U(1)$ symmetry.
The disceret $\mathbf{Z}_2$ symmetry represents the invariance of the system under the transformation
$x\to x+\lambda_c/2$ and $\phi_\alpha\to\phi_\alpha+\pi$.
The $U(1)$ symmetry represents the freedom of the total phase $\phi_\downarrow+\phi_\uparrow$ 
(i.e., the conservation of the total particle number): 
a simultaneous rotation of both condensate phases by an arbitrary constant phase, $\phi_\tau\to\phi_\tau+\varphi$, 
leaving $\Delta\phi$ invariant.
Note that the $U(1)$ symmetry associated with the freedom of the relative phase $\Delta\phi$
of the two condensate wavefunctions is nonetheless explicitly broken by the last term in Eq.~\eqref{E-functional-density}.
Considering solely this term, the minimization of the energy amounts to the constraint 
$\cos(k_cx)\cos{\phi_\alpha}\cos{\Delta\phi}<0$. This in turn imposes a position-dependent 
constraint on $\Delta\phi$, as the phase $\phi_\alpha$ of the cavity-field amplitude is constant over entire space.
Nonetheless, the kinetic-energy (i.e., first two) terms compete with this Raman coupling term and favors wavefunctions with uniform phases.

\emph{Self-ordering and symmetry breaking.}---We numerically solve equations~\eqref{eq:coupled-eqs}, 
assuming that the cavity-field amplitude quickly reaches its steady state,
\begin{align} \label{eq:ss-cavity-amplitude}
\alpha=\frac{\eta\int\cos(k_cx)
\left[\psi_\downarrow^*(x)\psi_{\uparrow}(x)+\psi_\uparrow^*(x)\psi_{\downarrow}(x)\right] dx}
{\Delta_c+i\kappa-\sum_\tau U_\tau \int \cos^2(k_cx)n_\tau(x)dx}.
\end{align}
One can identify
$\Theta\equiv\int\cos(k_cx)(\psi_\downarrow^*\psi_{\uparrow}+\text{H.c.})dx$ as an order parameter.
Unlike normal self-ordering in a single component BEC, here the cavity-field amplitude is coupled to 
the atomic spin polarization, rather than atomic density. That is, the scattering of pump-laser
photons into the cavity mode is accompanied by the atomic spin 
flip $\uparrow\leftrightarrow\downarrow$ and $\pm\hbar k_c$ momentum kick
along the $x$ direction. 

Using a self-consistent imaginary-time propagation method, we find 
the ground-state condensate wavefunctions
$\psi_\tau(x)$ and the steady-sate cavity-field amplitude $\alpha$.
Below a threshold pump strength $\eta_c$,
the cavity mode is empty and the condensate 
wavefunctions are uniform, with arbitrary phases. By increasing
the effective pump strength above the threshold $\eta_c$, 
the quantum fluctuations in the condensates trigger a constructive
scattering of pump-laser photons into the cavity mode via the two-photon Raman processes. 
These cavity photons in turn stimulate the Raman processes, 
leading into a random-field-induced runaway ordering process seeded by quantum fluctuations.

\begin{figure}[t!]
\centering
\includegraphics [width=0.48\textwidth]{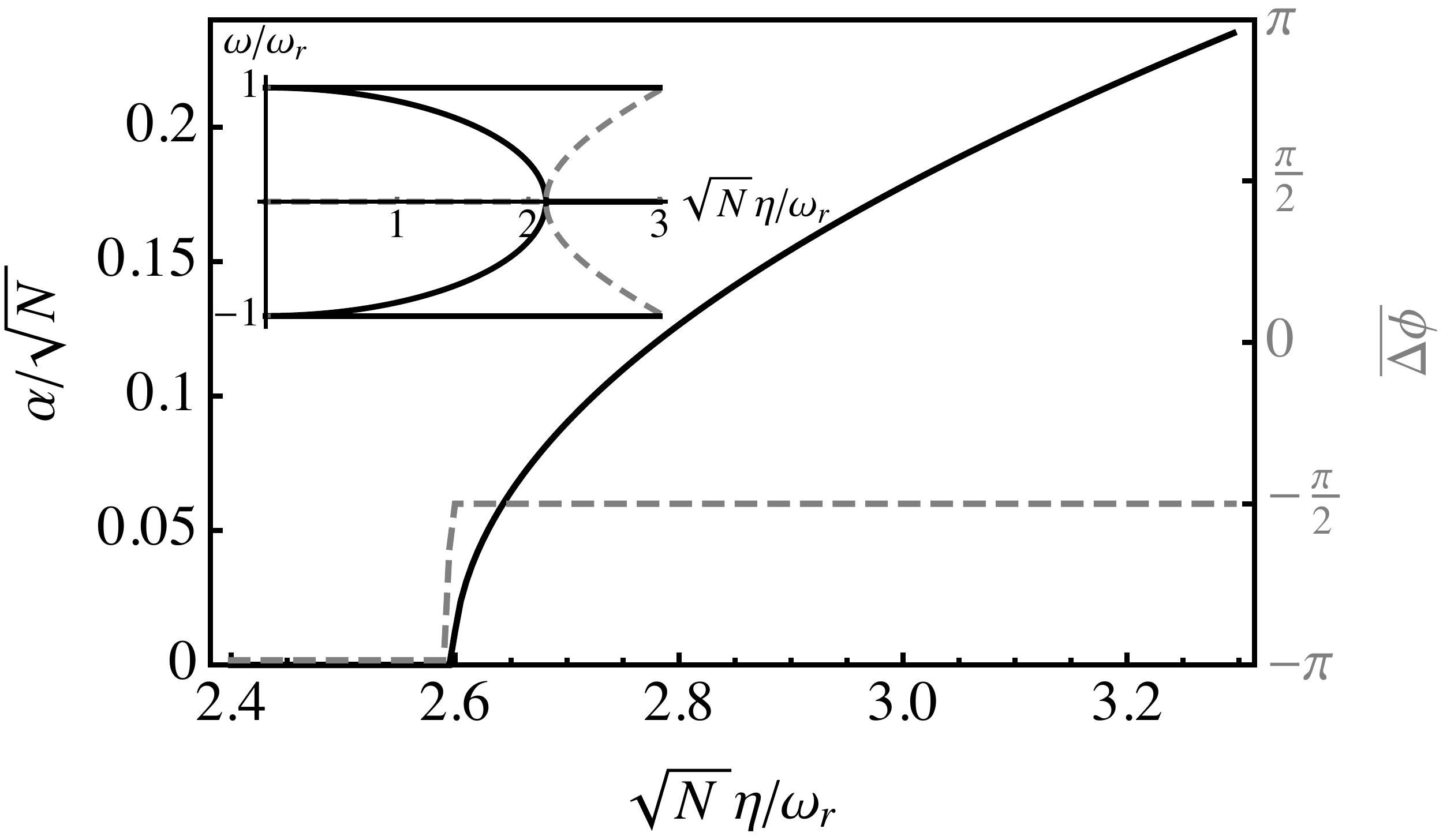}
\caption{Scaled steady-state cavity field amplitude as a function of the dimensionless transverse pump strength. 
The Dicke superradiance phase transition occurs around 
$\sqrt{N}\eta_c\approx2.6\omega_r$, where the cavity-field amplitude $\alpha$ 
(the black solid curve) attains a non-zero value and the average relative condensate 
phase $\overline{\Delta\phi}$ (the gray dashed curve) is locked at $-\pi/2$. 
The inset shows spectra of lowest four elementary 
excitations~\eqref{eq:characteristic-eq} exhibiting mode softening 
around $\sqrt{N}\eta_c\approx2.15\omega_r$, 
where the emergence of imaginary eigenvalues 
(the gray dashed curves) signals the superradiance phase transition.
The parameters are set to 
$(\Delta_c,NU_\downarrow,NU_\uparrow,\tilde{\delta},\kappa)=(-10,-2,-1,0.5,0)\omega_r$,
with the initial condition $f_\downarrow=0.8$ and $f_\uparrow=0.2$.}
\label{fig:alpha_relat-phase}
\end{figure}

\begin{figure}[t!]
\centering
\includegraphics [width=0.48\textwidth]{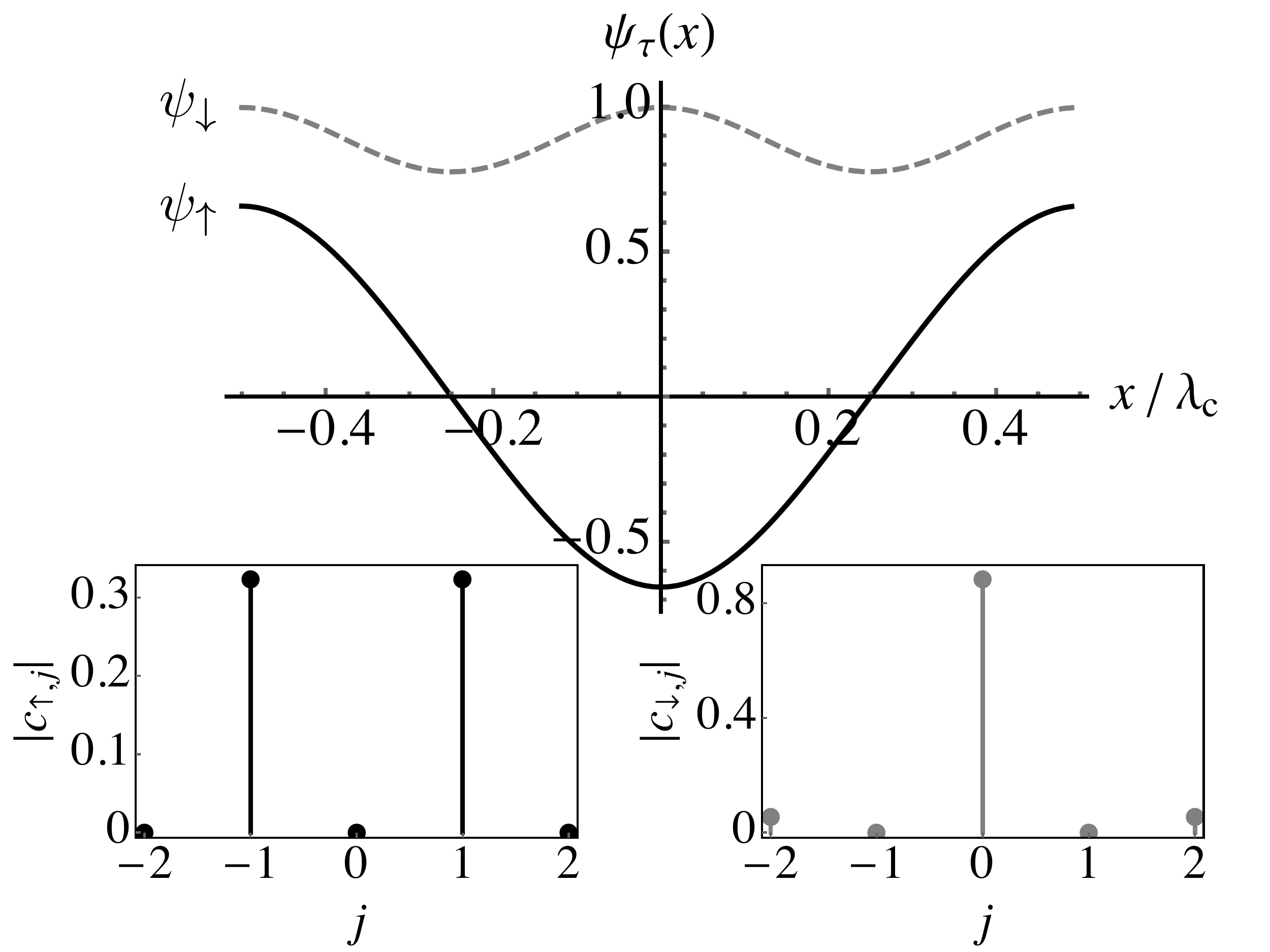}
\caption{The condensate wavefunctions $\psi_\tau(x)$ over one wavelength $\lambda_c$. 
The discrete $\lambda_c/2$-periodicity is broken by the spin-up wavefunction.
The relative condensate phase $\Delta\phi$ varies periodically in space 
with the average $\overline{\Delta\phi}=-\pi/2$
to minimize the total energy (for the current figure, $\phi_\alpha=0$).
The insets show the coefficients $|c_{\tau,j}|$ of the momentum-mode 
contributions to the condensate wavefunctions for a pump strength 
$\sqrt{N}\eta=3.2\omega_r$ and the other parameters as in Fig.~\ref{fig:alpha_relat-phase}.}
\label{fig:wavefunctions_fourier_coeffs}
\end{figure}

The results are presented in Figs.~\ref{fig:alpha_relat-phase} 
and ~\ref{fig:wavefunctions_fourier_coeffs} for the lossless cavity limit ($\kappa=0$). 
In Fig.~\ref{fig:alpha_relat-phase}, the solid black curve shows the scaled cavity-field amplitude $\alpha/\sqrt{N}$ as a function  
of the dimensionless effective pump strength $\sqrt{N}\eta/\omega_r$ for $\Delta_c=-10\omega_r$, 
$NU_\downarrow=-2\omega_r$, $NU_\uparrow=-\omega_r$, and 
$\tilde{\delta}=0.5\omega_r$, with $\omega_r=\hbar k_c^2/2m$ being the recoil frequency.
The initial condition for the fraction of atoms in each state is set to $f_\downarrow\equiv N_\downarrow/N=0.8$ and
$f_\uparrow\equiv N_\uparrow/N=0.2$ (see also the discussion on elementary excitations in the following).
For the given parameters, the superradiance phase transition occurs at the critical pump strength 
$\sqrt{N}\eta_c\approx2.6\omega_r$, where the symmetry between $\alpha$ and $-\alpha$ is spontaneously broken.

The corresponding condensate wavefunctions $\psi_\tau(x)$ deep in the superradiance phase are shown in 
Fig.~\ref{fig:wavefunctions_fourier_coeffs} for $\sqrt{N}\eta=3.2\omega_r$ and the 
other parameters as in Fig.~\ref{fig:alpha_relat-phase}. The black solid (gray dashed) 
curve represents $\psi_\uparrow$ ($\psi_\downarrow$).
Although the spin-down condensate wavefunction $\psi_\downarrow$
is $\lambda_c/2$ periodic, the spin-up condensate wavefunction  
$\psi_\uparrow$ breaks this discrete $\lambda_c/2$-translational symmetry.
Note that the condensate densities $n_\tau(x)=|\psi_\tau(x)|^2$ for both 
components clearly have two identical peaks within one $\lambda_c$ and 
are, therefore, $\lambda_c/2$-periodic~\cite{footenote1}.
Therefore, the $\mathbf{Z}_2$ symmetry is broken by a spin wave, 
rotating along the $y$ axis as 
$\psi_{x,\uparrow}(-\lambda_c/2)\equiv\psi_{\downarrow}^{\rm max}+|\psi_{\uparrow}^{\rm max}|$,
$\psi_{\downarrow}^{\rm min}(-\lambda_c/4)$, 
$\psi_{x,\downarrow}(0)\equiv\psi_{\downarrow}^{\rm max}-|\psi_{\uparrow}^{\rm max}|$, 
and $\psi_{\downarrow}^{\rm min}(\lambda_c/4)$, 
rather than the density ordering.
This is in a sharp contrast to the normal one-component
self-ordered BEC, where the condensate density is $\lambda_c$-periodic in order
to allow for in-phase constructive scattering of the pump-laser photons into the
cavity mode. This is not the case in our model as the cavity field is coupled to 
the atomic spin polarization rather than density [see Eq.~\eqref{eq:ss-cavity-amplitude}],
and it will be discussed in more detail in the following. 
Note also that the $U(1)$ symmetry representing the freedom of the total phase
is broken here by imposing initial conditions in numerical calculations, 
resulting in purely real condensate wavefunctions for both components.

In the self-ordered state, the relative condensate phase $\Delta\phi(x)$ varies in space. For $\alpha>0$, 
the phase of the cavity-field amplitude is fixed over the entire space: $\phi_\alpha=0$.
In order to satisfy the constraint $\cos(k_cx)\cos\phi_\alpha\cos{\Delta\phi}<0$, 
the relative condensate phase must therefore be 
$\Delta\phi=\pm\pi$ (modulo $2\pi$) for $|x|\leqslant\lambda_c/4$ (e.g., for real wavefunctions 
one condensate wavefunction is positive and the other negative) and $\Delta\phi=0$ (modulo $2\pi$) 
for $\lambda_c/4<|x|\leqslant\lambda_c/2$ (e.g., for real wavefunctions both condensate 
wavefunctions are either positive or negative);
see Fig.~\ref{fig:wavefunctions_fourier_coeffs}. 
For $\alpha<0$ (i.e., $\phi_\alpha=\pi$), this is opposite, namely
$\Delta\phi=0$ (modulo $2\pi$) for $|x|\leqslant\lambda_c/4$ and $\Delta\phi=\pm\pi$ (modulo $2\pi$) 
for $\lambda_c/4<|x|\leqslant\lambda_c/2$. 
This constraint on the relative condensate phase due to the energy is also consistent
with Eq.~\eqref{eq:ss-cavity-amplitude}. 
Consider for instance the $\alpha>0$ case, which implies that 
$\Delta\phi=\pm\pi$ for $|x|\leqslant\lambda_c/4$ and $\Delta\phi=0$
for $\lambda_c/4<|x|\leqslant\lambda_c/2$. This in turn ensures that
the order parameter $\Theta$ is non-zero and negative: for $|x|\leqslant\lambda_c/4$
one has $\cos(k_cx)>0$ and $\cos{\Delta\phi}<0$,
while $\cos(k_cx)<0$ and $\cos{\Delta\phi}>0$ for $\lambda_c/4<|x|\leqslant\lambda_c/2$.
Since $\Delta_c-\sum_\tau \int \cos^2(k_cx)U_\tau n_\tau(x)dx$ must be negative to avoid heating, 
this yields a positive cavity-field amplitude $\alpha>0$ in a self-consistent way.

Around the threshold $\eta_c$ where $|\alpha|/\sqrt{N}\ll1$,  
the weakly Raman-coupled two-component BEC 
is equivalent to the two-dimensional classical ferromagnetic $XY$ spin model in a uniaxial random magnetic field, 
where the relative condensate phase $\Delta\phi$ plays the role of the spin angle
and the position-dependent Raman field with the zero spatial average  
mimics the random magnetic field~\cite{Niederberger2008,SM-Mivehvar2017}.
The classical $XY$ spin model in two dimension has no net 
magnetization in the absence of the magnetic field due to 
the Mermin-Wagner-Hohenberg (MWH) no-go theorem~\cite{Mermin1966,Hohenberg1967},
while a weak uniaxial random magnetic field breaks the continuos $U(1)$ symmetry
associated with the freedom of spin angles and hence violates the applicability conditions of the MWH theorem,
resulting in a spontaneous magnetization perpendicular to the random field~\cite{Feldman1998}, even in small finite temperatures~\cite{Minchau1985,Wehr2006}.
The applicability of the MWH theorem is also violated in our system by the explicitly broken $U(1)$ symmetry
via the Raman field, as well as cavity-mediated long range interactions~\cite{SM-Mivehvar2017}.
The corresponding zero-average-Raman-field induced order in the two-component BEC manifests in
the averaged relative condensate phase 
$\overline{\Delta\phi}=(1/\lambda_c)\int_{0}^{\lambda_c}\Delta\phi(x) dx$, 
which is locked at either $\pi/2$ or $-\pi/2$~\cite{Niederberger2008}.
The averaged relative condensate phase $\overline{\Delta\phi}$
is shown in Fig.~\ref{fig:alpha_relat-phase} as the gray dashed curve.
It takes the value $-\pi/2$ (which is favored over $\pi/2$ in our calculations duo to initial conditions) 
in the self-ordered state and an arbitrary value ($\approx-\pi$) otherwise.
Note that even for exceedingly small Raman couplings $\eta|\alpha|\ll\omega_r$
in the onset of the Dicke superradiance, $\overline{\Delta\phi}$ is still fixed at $-\pi/2$, 
indicating that the superradiance phase transition in our model is indeed a random-field induced
ordering process similar to the two-dimensional classical $XY$ spin model in a uniaxial random magnetic field.
Nevertheless, the seed of the random Raman field stems form quantum fluctuations
and it is a dynamical entity: the random Raman field triggers the self-ordering of the BEC and 
this in turn amplifies the dynamical random Raman field, resulting in a runaway process. 

For $\tilde{\delta}>0$ and red-detuned cavity optical potentials 
$\hbar U_\tau|\alpha|^2\cos^2(k_cx)<0$ with $|U_\downarrow|>|U_\uparrow|$, 
the spin-down condensate is energetically favored over the spin-up component: 
$|\psi_\downarrow(x)|\geqslant|\psi_\uparrow(x)|$.
Due to the kinetic energy cost, the spin-up condensate wavefunction $\psi_\uparrow(x)$ 
is therefore favored to change its phase $\phi_\uparrow(x)$ over space at $x=(l+1/2)\lambda_c/2$
(with $l\in\mathbb{Z}$) to fulfill the phase constraint discussed above;
see Fig.~\ref{fig:wavefunctions_fourier_coeffs}. This implies that the zero-momentum mode
of the spin-up condensate must be depleted. Due to the discrete $\lambda_c$-translation
symmetry, one can decompose the condensate wavefunctions into traveling plane waves as
$\psi_\tau(x)=\sum_jc_{\tau,j}e^{ijk_cx}$. The coefficients $c_{\tau,j}$ of the lowest three 
momentum modes $j=0,\pm1,\pm2$ for both condensates are depicted in the insets of 
Fig.~\ref{fig:wavefunctions_fourier_coeffs}. As expected, for the spin-up condensate 
the zero-momentum mode $j=0$ is almost completely depleted and the first excited momentum
states $j=\pm1$ are highly populated, while for the spin-down condensate 
the zero-momentum mode has the dominant population with a small contribution from 
the second excited momentum states $j=\pm2$. Such a coupling of different spins to different momentum states
is reminiscent of a synthetic spin-orbit interaction~\cite{Lin2011}. 
All higher momentum modes have a negligible population for both condensates.

The equations of motion~\eqref{eq:coupled-eqs} can be linearized for 
small quantum fluctuations around the mean-field solutions to yield 
spectrum $\omega$ of elementary excitations~\cite{Horak2001,Nagy2008}. To this end, we linearize
Eq.~\eqref{eq:coupled-eqs} around the trivial solution $\alpha=0$ and $\psi_\tau=\sqrt{Nf_\tau}$,
with restricting atomic excitations to solely $e^{\pm ik_cx}$ owing to the Raman coupling
(or cavity pump) $\propto\cos(k_cx)$, and obtain a sixth order
characteristic equation~\cite{SM-Mivehvar2017}, 
\begin{align} \label{eq:characteristic-eq}
(\omega^2-\omega_r^2)\left\{(\omega^2-\omega_r^2)\left[\delta_c^2+(i\omega-\kappa)^2\right]
+2\delta_c\omega_rN\eta^2\right\}=0,
\end{align}
with $\delta_c\equiv-\Delta_c+N\sum_\tau f_\tau U_\tau/2$.
The lowest four spectra composed mainly of the atomic excitations are shown 
in the inset of Fig.~\ref{fig:alpha_relat-phase} as a function of the dimensionless pump strength. 
Two of the excitation branches, corresponding mainly to the atomic excitations of the spin-up component, approach
zero by increasing $\eta$ and at the same time a pair of imaginary eigenvalues (dashed curves) appear, signaling dynamical
instability in the trivial solution. This is the onset of the superradiance phase transition.
The zero frequency $\omega=0$ solution of the characteristic equation~\eqref{eq:characteristic-eq} 
yields the threshold pump value,
\begin{align}
\sqrt{N}\eta_c=\sqrt{\frac{\delta_c^2+\kappa^2}{2\delta_c}}\sqrt{\omega_r}\approx2.15\omega_r.
\end{align}  
This is smaller than the threshold $\sqrt{N}\eta_c\approx2.6\omega_r$ obtained
from the mean-filed calculation. This can be attributed to the fact that here the atomic excitations are
restricted to the lowest manifold $e^{\pm ik_cx}$, while this is not strictly true as can be seen from the insets of 
Fig.~\ref{fig:wavefunctions_fourier_coeffs}. 

Finally, let us briefly comment on two simplifying assumptions used throughout the paper. First, adding weak two-body interactions will somewhat flatten the condensate wavefunctions without qualitative change of the results. However, strong two-body interactions along with a Raman coupling can give rise to a state with out-phase density modulations
of the two Bose condensates~\cite{Abad2013}. Second, a non-zero cavity decay $\kappa>0$ results in an imaginary $\alpha$ with a phase $\phi_\alpha\in[0,2\pi]$, rather than 0 or $\pi$ as discussed above for $\kappa=0$. Nonetheless, the averaged relative condensate phase $\overline{\Delta\phi}$ would be still locked at $\pm\pi/2$ as before.
A finite decay can also induce cavity-mediated collisional relaxation. 
However, these collisions are ineffective up to long time scales 
proportional to the atom number $N$~\cite{piazza_QKE}.

\emph{Conclusions.}---We studied spatial spin and density self-ordering of ultracold bosonic atoms 
with two internal ground states coupled via a single cavity mode. 
Interestingly, spin and density self-orderings are closely tied together due to 
light-induced spin-orbit coupling, which emerges as a result of an order-by-disorder 
process induced by a cavity-assisted Raman field.  
Although our model constitutes the simplest example of multi-component itinerant atoms coupled to
dynamical cavity fields, it already highlights the rich physics which
can result from the strong
photon-induced spin- and position-dependent interactions between atoms in these
systems (a related idea has been recently put forward for atoms
coupled to photonic crystal waveguides~\cite{Manzoni2017}). 
Therefore, our work opens a new avenue for exploring a wealth of novel many-body phenomena, including spin-Peierls transition, topological insulators, interaction-driven fractional topological and symmetry-broken phases, etc., in tunable multi-component cavity-QED environments.



We are grateful to Alessio Recati, Tobias Donner, and Manuele Landini for fruitful discussions.
FM and HR are supported by the Austrian Science Fund project I1697-N27.

\bibliography{2BEC_lin_cav}

\begin{thebibliography}{50}
\bibitem [{foo()}]{footenote1}%
  \BibitemOpen
\bibfield  {journal} {  }\href@noop {} {\bibinfo  {journal} {For
  $U_\downarrow=U_\uparrow$ and $\tilde{\delta}=0$, the ground state is doubly
  degenerate and the condensate densities are $\lambda_c$ periodic. One ground
  state corresponds to both condensates having density peaks in even sites and
  the other corresponds to both condensates having density peaks in odd sites}\
  }\BibitemShut {NoStop}%
\end{thebibliography}

\newpage
\widetext
\setcounter{equation}{0}
\renewcommand{\theequation}{S\arabic{equation}}

\section{Supplemental Material}
Here, we present the details of the adiabatic elimination of the atomic excited states,
equivalence of the Raman-coupled two-component Bose-Einstein condensate (BEC) and the two-dimensional (2D) classical $XY$ spin model in a uniaxial magnetic field, 
and derivation of the elementary-excitation spectra of the system.

\subsection{Adiabatic Elimination of the Atomic Excited States}

Consider four-level bosonic atoms inside a linear cavity, which are illuminated in the transverse direction by two external standing-wave pump lasers as depicted in Fig.~1 in the main text. A tight confinement along the transverse directions has been assumed so that the atomic motion along these directions is frozen. The transition $\uparrow\>\leftrightarrow1$ ($\downarrow\>\leftrightarrow2$) is induced by the first (second) external transverse laser with the Rabi frequency $\Omega_1$ ($\Omega_2$), while the transitions $\downarrow\>\leftrightarrow 1$ and $\uparrow\>\leftrightarrow 2$ are coupled to a cavity mode with identical coupling strength $\mathscr{G}(x)=\mathscr{G}_0\cos(k_cx)$ which is initially in the vacuum state. The scheme basically constitutes  a double $\Lambda$ configuration, where $\{\ket{\downarrow},\ket{\uparrow}\}$ are the ground pseudo-spin states and $\{\ket{1},\ket{2}\}$ are electronic excited states, with energies $\{\hbar\omega_\downarrow=0, \hbar\omega_\uparrow,\hbar\omega_{1},\hbar\omega_{2}\}$. The pump and cavity frequencies, respectively, $\{\omega_{p1},\omega_{p2}\}$ and $\omega_c$ are assumed to be far red detuned from the atomic transition frequencies.

The single-particle Hamiltonian density in the dipole and the rotating wave approximations reads:
\begin{align} \label{eq:1-particle-H-density}
\mathcal{H}^{(1)}=\frac{p^2}{2m}I_{4\times4}
+\sum_{\tau=\{\uparrow,1,2\}}\hbar\omega_\tau\sigma_{\tau \tau}
+\hbar\omega_c\hat{a}^\dagger\hat{a}
+\hbar\Big[\mathscr{G}(x)\hat{a}(\sigma_{1\downarrow}+\sigma_{2\uparrow}) 
+\Omega_1e^{-i\omega_{p1}t}\sigma_{1\uparrow}
+\Omega_2e^{-i\omega_{p2}t}\sigma_{2\downarrow}
+ \text{H.c.}\Big],
\end{align}
where $m$ is the atomic mass, $\hat{a}$ is the annihilation operator of cavity photons, $\sigma_{\tau \tau'}=\ket{\tau}\bra{\tau'}$, and H.c.\ stands for the Hermitian conjugate. Here, $p$ is the center-of-mass momentum operator of the atom along the cavity axis $x$, and and $I_{4 \times 4}$ is the identity matrix in the internal atomic-state space. The single-particle Hamiltonian density~\eqref{eq:1-particle-H-density} can be conveniently expressed in the rotating frame of the pump laser 
\begin{align} \label{eq:1-particle-H-density-rot-frame}
\tilde{\mathcal{H}}^{(1)}&=\frac{p^2}{2m}I_{4\times4}
+\hbar\delta\sigma_{\uparrow\uparrow}
-\hbar\Delta_1\sigma_{11}-\hbar\Delta_2\sigma_{22}
-\hbar\Delta_c\hat{a}^\dagger\hat{a}
+\hbar\Big[\mathscr{G}(x)\hat{a}(\sigma_{1\downarrow}+\sigma_{2\uparrow}) 
+\Omega_1\sigma_{1\uparrow}+\Omega_2\sigma_{2\downarrow}
+ \text{H.c.}\Big],
\end{align}
through $\tilde{\mathcal{H}}^{(1)}=\mathscr{U}\mathcal{H}^{(1)}\mathscr{U}^\dag+i\hbar(\partial_t \mathscr{U})\mathscr{U}^\dag$ 
and exploiting the unitary transformation
\begin{align} \label{U-T}
\mathscr{U}=\exp{\left\{i\left[\left(\frac{\omega_{p1}+\omega_{p2}}{2}\right)\hat{a}^\dagger\hat{a}
+\left(\frac{\omega_{p2}-\omega_{p1}}{2}\right)\sigma_{\uparrow\uparrow}
+\left(\frac{\omega_{p1}+\omega_{p2}}{2}\right)\sigma_{11}+\omega_{p2}\sigma_{22}\right]t\right\}}.
\end{align}
Here, we have defined 
$\Delta_1\equiv(\omega_{p1}+\omega_{p2})/2-\omega_1$,
$\Delta_2\equiv\omega_{p2}-\omega_2$,
and $\Delta_c\equiv(\omega_{p1}+\omega_{p2})/2-\omega_c$ as the atomic and cavity detunings 
with respect to the pump lasers, respectively, and $\delta\equiv\omega_\uparrow-(\omega_{p2}-\omega_{p1})/2$
is the relative two-photon detuning.

The corresponding many-body Hamiltonian takes the form,
\begin{align} \label{eq:many-body-H}
H=\int \hat\Psi^\dag(x)\tilde{\mathcal{M}}^{(1)}
\hat\Psi(x)dx+H_{\text{int}}, 
\end{align}
where $\hat\Psi(x)=(\hat\psi_\downarrow,\hat\psi_\uparrow,\hat\psi_1,\hat\psi_2)^{\mathsf{T}}$ are the bosonic atomic field operators 
satisfying the usual bosonic commutation relation 
$[\hat\psi_\tau(x),\hat\psi_{\tau'}^\dag(x')]=\delta_{\tau,\tau'}\delta(x-x')$, 
and $\tilde{\mathcal{M}}^{(1)}$ is the matrix form of the Hamiltonian density $\tilde{\mathcal{H}}^{(1)}$, 
Eq.~\eqref{eq:1-particle-H-density-rot-frame}.
The interaction Hamiltonian $H_{\text{int}}$ accounts for the two-body contact interactions between atoms
and will be omitted in the following.
The Heisenberg equations of motion of the photonic and atomic field operators can be obtained using 
the many-body Hamiltonian~\eqref{eq:many-body-H},
\begin{align} \label{eq:Heisenberg-eqs-motion}
i\hbar\frac{\partial}{\partial t}\hat{a}&=-\hbar\Delta_c\hat{a}-i\hbar\kappa\hat{a}
+\hbar\int \mathscr{G}(x)\left[\hat\psi_\downarrow^\dag(x)\hat\psi_{1}(x)+\hat\psi_\uparrow^\dag(x)\hat\psi_{2}(x)\right] dx,\nonumber\\
i\hbar\frac{\partial}{\partial t}\hat\psi_\downarrow&=-\frac{\hbar^2}{2m}\frac{\partial^2}{\partial x^2}\hat\psi_\downarrow
+\hbar\mathscr{G}(x)\hat{a}^\dag\hat\psi_1+\hbar\Omega_2\hat\psi_2,\nonumber\\
i\hbar\frac{\partial}{\partial t}\hat\psi_\uparrow&=\left(-\frac{\hbar^2}{2m}\frac{\partial^2}{\partial x^2}+\hbar\delta\right)\hat\psi_\uparrow
+\hbar\mathscr{G}(x)\hat{a}^\dag\hat\psi_2+\hbar\Omega_1\hat\psi_1,\nonumber\\
i\hbar\frac{\partial}{\partial t}\hat\psi_1&=\left(-\frac{\hbar^2}{2m}\frac{\partial^2}{\partial x^2}-\hbar\Delta_1\right)\hat\psi_1
+\hbar\mathscr{G}(x)\hat{a}\hat\psi_\downarrow+\hbar\Omega_1\hat\psi_\uparrow,\nonumber\\
i\hbar\frac{\partial}{\partial t}\hat\psi_2&=\left(-\frac{\hbar^2}{2m}\frac{\partial^2}{\partial x^2}-\hbar\Delta_2\right)\hat\psi_2
+\hbar\mathscr{G}(x)\hat{a}\hat\psi_\uparrow+\hbar\Omega_2\hat\psi_\downarrow,
\end{align}
where we have phenomenologically included the decay of the cavity mode $-i\hbar\kappa\hat{a}$. 

Let us now assume that atomic detunings $\Delta_{j}$ are large so that the atomic field operators
$\{\hat\psi_1,\hat\psi_2\}$ of the excited states reach steady states very fast and we can therefore adiabatically 
eliminate their dynamics. Omitting the kinetic energies in comparison to $-\hbar\Delta_1$ and $-\hbar\Delta_2$, 
we obtain the steady-state atomic field operators of the excited states,
\begin{align} \label{eq:ss-atomic-excited-field-op}
\hat\psi_1^{\text{ss}}=\frac{1}{\Delta_1}
\left[\mathscr{G}(x)\hat{a}\hat\psi_\downarrow+\Omega_1\hat\psi_\uparrow\right],\nonumber\\
\hat\psi_2^{\text{ss}}=\frac{1}{\Delta_2}
\left[\mathscr{G}(x)\hat{a}\hat\psi_\uparrow+\Omega_2\hat\psi_\downarrow\right].
\end{align}  
Substituting the steady-state atomic field operators of the excited states~\eqref{eq:ss-atomic-excited-field-op} 
in the Heisenberg equations of motion~\eqref{eq:Heisenberg-eqs-motion} yields a set of effective equations for the photonic and ground-state atomic field operators,
\begin{align}  \label{eq:eff-Heisenberg-eqs-motion}
i\hbar\frac{\partial}{\partial t}\hat{a}&=\hbar\left\{-\Delta_c-i\kappa
+\int \cos^2(k_cx)
\left[U_\downarrow\hat\psi_\downarrow^\dag(x)\hat\psi_\downarrow(x)
+U_\uparrow\hat\psi_\uparrow^\dag(x)\hat\psi_\uparrow(x)\right] dx\right\}\hat{a}\nonumber\\
&+\hbar\eta\int\cos(k_cx)\left[\hat\psi_\uparrow^\dag(x)\hat\psi_\downarrow(x)
+\hat\psi_\downarrow^\dag(x)\hat\psi_\uparrow(x)\right] dx,\nonumber\\
i\hbar\frac{\partial}{\partial t}\hat\psi_\downarrow&=\left[-\frac{\hbar^2}{2m}\frac{\partial^2}{\partial x^2}
+\hbar U_\downarrow\hat{a}^\dag\hat{a}\cos^2(k_cx)\right]\hat\psi_\downarrow
+\hbar\eta\cos(k_cx)(\hat{a}^\dag+\hat{a})\hat\psi_\uparrow,\nonumber\\
i\hbar\frac{\partial}{\partial t}\hat\psi_\uparrow&=\left[-\frac{\hbar^2}{2m}\frac{\partial^2}{\partial x^2}
+\hbar\tilde{\delta}+\hbar U_\uparrow\hat{a}^\dag\hat{a}\cos^2(k_cx)\right]\hat\psi_\uparrow
+\hbar\eta\cos(k_cx)(\hat{a}^\dag+\hat{a})\hat\psi_\downarrow,
\end{align}
where we have introduced 
$\tilde{\delta}\equiv\delta+\Omega_1^2/\Delta_1-\Omega_2^2/\Delta_2$,
$U_{\downarrow(\uparrow)}\equiv\mathscr{G}_0^2/\Delta_{1(2)}$, 
and $\eta\equiv\mathscr{G}_0\Omega_1/\Delta_1=\mathscr{G}_0\Omega_2/\Delta_2$.
Replacing the photonic and atomic field operators $\hat{a}$ and 
$\hat\psi_\tau$ with their corresponding averages $\alpha\equiv\langle\hat{a}\rangle$ 
and $\psi_\tau\equiv\langle\hat\psi_\tau\rangle$, respectively, yields Eq.~(1) in the main text.

\section{Equivalence of the Raman-coupled two-component BEC and the 2D classical $XY$ model in a uniaxial magnetic field}

Without loss of generality, consider equal densities $n_\uparrow=n_\downarrow=n=\text{const.}$ for the two condensate components in the absence of the cavity field.
In the onset of the superradiant state, we have
$|\alpha|/\sqrt{N}\ll1$. Therefore, one can assume that the condensate densities are not significantly different from each other
and one still has $n_\uparrow\simeq n_\downarrow\simeq n$~\cite{Niederberger2008}.
Using these assumptions, the energy-functional density, Eq.~(2) in the manuscript, around the superradiant phase transition can be recast as follows:
\begin{align} \label{eq:SM-E-density-recast}
\mathscr{E}_{\rm PT}(x)\simeq \frac{\hbar^2}{m}(\partial_x\sqrt{n})^2
+\frac{\hbar^2}{4m}n\left[(\partial_x\Delta\phi)^2+(\partial_x\Phi)^2\right]
+\hbar\tilde{\delta}n
+4\hbar\eta|\alpha|n\cos(k_cx)\cos{\phi_\alpha}\cos{\Delta\phi},
\end{align}
where $\Phi\equiv\phi_\downarrow+\phi_\uparrow$.
Note that the terms proportional to $|\alpha|^2$
have been omitted as  $|\alpha|^2/N\ll|\alpha|/\sqrt{N}\ll1$
(keeping them does not change the following discussion in anyway).
Minimization of the energy ensures that $\Phi=\text{const.}$
The energy-functional density \eqref{eq:SM-E-density-recast} then reduces to
\begin{align} \label{eq:SM-E-density-recast-simplified}
\mathscr{E}_{\rm PT}(x)&=
\frac{\hbar^2}{m}(\partial_x\sqrt{n})^2+\hbar\tilde{\delta}n
+\frac{\hbar^2}{4m}n(\partial_x\Delta\phi)^2
+4\hbar\eta n|\alpha|\cos{\phi_\alpha}\cos(k_cx)\cos{\Delta\phi},
\end{align}
where the first two terms in Eq.~\eqref{eq:SM-E-density-recast-simplified}
play no role in determining the relative condensate phase $\Delta\phi$. 

The 2D classical $XY$ spin model in a magnetic field $\mathbf{h}_i$ is described by the Hamiltonian,
\begin{align} \label{eq:SM-XY-H}
H_{XY}=-J\sum_{\langle i,j\rangle} \mathbf{s}_i\cdot\mathbf{s}_j+\sum_i \mathbf{h}_i\cdot\mathbf{s}_i,
\end{align}
where ${\langle i,j\rangle}$ indicates that the sum is over nearest neighbors.
Parametrizing the spins with their angles $\theta_i$ with respect to the $x$ axis as 
$\mathbf{s}_i=(\cos\theta_i,\sin\theta_i,0)$ and assuming a uniaxial random magnetic field $\mathbf{h}_i=(h_i,0,0)$ with $\sum_ih_i=0$~\cite{Wehr2006},
Eq.~\eqref{eq:SM-XY-H} can be recast as,
\begin{align} \label{eq:SM-XY-H-parametrized}
H_{XY}=-J\sum_{\langle i,j\rangle}\cos{(\theta_i-\theta_j)}+\sum_i h_i\cos\theta_i.
\end{align}
Assuming small spin-angle fluctuations around a possible ordered state and going to the continuum limit,
one obtains
\begin{align} \label{eq:SM-XY-H-parametrized-continuum}
H_{XY}^{\rm cont}\simeq \int dx dy \left[-J(\nabla\theta)^2+h(x,y)\cos\theta\right].
\end{align}
One can now immediately see the resemblance between the 2D classical $XY$ spin Hamiltonian $H_{XY}^{\rm cont}$ in the uniaxial random magnetic field in the
continuum limit and the energy functional of the cavity-assisted Raman-coupled two-component BEC around the Dicke superradiance phase transition
$E_{\rm PT}=\int \mathscr{E}_{\rm PT}(x)dx$. The relative condensate phase $\Delta\phi$ plays the role of the spin angle $\theta$
and the position-dependent Raman field $\propto \eta|\alpha|\cos\phi_\alpha\cos(k_cx)$ with the zero spatial average mimics the random magnetic field 
$h(x,y)$~\cite{Niederberger2008}.

The Mermin-Wagner-Hohenberg (MWH) no-go theorem states that 
continuous symmetries cannot be spontaneously broken at finite temperature in systems with sufficiently short-range interactions in 
1D and 2D. The uniaxial random magnetic field in the 2D classical $XY$ spin model \eqref{eq:SM-XY-H-parametrized} explicitly breaks the rotational symmetry
of the system and the MWH theorem is, therefore, no longer applicable. Hence, the system magnetizes in the transverse $y$ 
direction due to the random magnetic field~\cite{Feldman1998}, even at small non-zero temperatures~\cite{Wehr2006,Minchau1985}.

On the other hand, our system \eqref{eq:SM-E-density-recast-simplified} violates the conditions of the MWH theorem in two ways: 
i)~by explicitly breaking the continuous $U(1)$ symmetry via the zero-average Raman field~\cite{Niederberger2008}, 
as in the 2D classical $XY$ spin model in the uniaxial random magnetic field~\cite{Wehr2006,Feldman1998,Minchau1985}, 
and ii)~by cavity-induced long-range interactions. The latter are present because
the Raman field is dynamic, i.e., it adapts to the atomic configuration.
Since the MWH theorem is not applicable
to our system as well, the relative condensate phase locked at $\pm\pi/2$ due
to the cavity-induced zero-average Raman field should persist 
even for small non-zero temperatures.

Let us briefly comment on the thermalization issue.
Since our system is driven-dissipative, no notion of temperature is in
general available in the steady state as the relaxation does not
lead to a thermal state~\cite{piazza_QKE}. 
However, due to the long-range nature of cavity-mediated interactions, the
steady-state is never reached in the thermodynamic limit
$N,V\to\infty$ and $N/V\to\text{const.}$, since the relaxation time is extensive: $t_{\rm relax}\propto N$.
Therefore, in the thermodynamic limit and for a finite time, the atoms do not experience any
relaxation and remain in a thermal distribution in which they were
prepared initially before being coupled to the cavity field. 
At relevant experimental time scales, the
only effect of the cavity is, therefore, to provide a dynamical Raman field. In
particular, at low enough temperatures where we can consider only the
BEC components of the atoms, our system is described by Eq.~(1) in the main manuscript.

\subsection{Elementary Excitations}

Assuming $\psi_\tau=\psi_{0\tau}+\delta\psi_\tau$ and $\alpha=\alpha_0+\delta\alpha$~\cite{Horak2001,Nagy2008},
where $\delta\psi_\tau$ and $\delta\alpha$ are quantum fluctuations around the mean-field
solutions $\psi_{0\tau}$ and $\alpha_0$, we linearize Eq.~(1) in the main text:
\begin{align} \label{eq:SM_lin_eqs}
i\frac{\partial}{\partial t}\delta\alpha&=\Big[-\Delta_c-i\kappa
+\sum_{\tau=\downarrow,\uparrow}U_\tau \int \cos^2(k_cx) n_{0\tau} dx\Big]\delta\alpha
+\alpha_0\sum_{\tau=\downarrow,\uparrow}U_\tau\int \cos^2(k_cx)
(\psi_{0\tau}\delta\psi_\tau^*+\psi_{0\tau}^*\delta\psi_\tau) dx\nonumber\\
&+\eta\int\cos(k_cx)
\left(\psi_{0\uparrow}\delta\psi_\downarrow^*+\psi_{0\downarrow}^*\delta\psi_\uparrow
+\psi_{0\uparrow}^*\delta\psi_\downarrow+\psi_{0\downarrow}\delta\psi_\uparrow^*\right) dx,
\nonumber\\
i\frac{\partial}{\partial t}\delta\psi_\downarrow&=
\Big[-\frac{\hbar}{2m}\frac{\partial^2}{\partial x^2}
+U_\downarrow|\alpha_0|^2\cos^2(k_cx)\Big]\delta\psi_\downarrow
+\eta(\alpha_0+\alpha_0^*)\cos(k_cx)\delta\psi_\uparrow
+U_\downarrow\cos^2(k_cx)\psi_{0\downarrow}(\alpha_0\delta\alpha^*+\alpha_0^*\delta\alpha)\nonumber\\
&+\eta\cos(k_cx)\psi_{0\uparrow}(\delta\alpha+\delta\alpha^*),
\nonumber\\
i\frac{\partial}{\partial t}\delta\psi_\uparrow&=
\Big[-\frac{\hbar}{2m}\frac{\partial^2}{\partial x^2}+\tilde{\delta}
+U_\uparrow|\alpha_0|^2\cos^2(k_cx)\Big]\delta\psi_\uparrow
+\eta(\alpha_0+\alpha_0^*)\cos(k_cx)\delta\psi_\downarrow
+U_\uparrow\cos^2(k_cx)\psi_{0\uparrow}(\alpha_0\delta\alpha^*+\alpha_0^*\delta\alpha)\nonumber\\
&+\eta\cos(k_cx)\psi_{0\downarrow}(\delta\alpha+\delta\alpha^*).
\end{align}
Assuming ans\"{a}tze 
$\delta\psi_\tau=e^{-i\mu_\tau t/\hbar}(\delta\psi_{+\tau}e^{-i\omega t}+\delta\psi_{-\tau}^*e^{i\omega t})$
and $\delta\alpha=\delta\alpha_{+}e^{-i\omega t}+\delta\alpha_{-}^*e^{i\omega t}$ for the quantum fluctuations, 
where the chemical potentials for the two components are equal $\mu_1=\mu_2$ when $\alpha_0\neq0$, one
can recast Eq.~\eqref{eq:SM_lin_eqs} in a matrix form for the positive and negative components,
\begin{align}
\omega
\begin{pmatrix}
\delta\alpha_+ \\
\delta\alpha_- \\
\delta\psi_{+\downarrow} \\
\delta\psi_{-\downarrow} \\
\delta\psi_{+\uparrow} \\
\delta\psi_{-\uparrow}
\end{pmatrix}
=\mathbf{M}_{\rm B}
\begin{pmatrix}
\delta\alpha_+ \\
\delta\alpha_- \\
\delta\psi_{+\downarrow} \\
\delta\psi_{-\downarrow} \\
\delta\psi_{+\uparrow} \\
\delta\psi_{-\uparrow}
\end{pmatrix},
\end{align}
where $\mathbf{M}_{\rm B}$ is a non-Hermitian matrix determined straightforwardly from Eq.~\eqref{eq:SM_lin_eqs}.
The eigenvalues of the matrix $\mathbf{M}_{\rm B}$ gives Bogoliubov-type excitations of the condensates 
and field. The existence of imaginary eigenvalues signals dynamical instability in the mean-field solutions. 
Analyzing the stability of uniform condensate wavefunctions and vacuum cavity field $\alpha_0=0$ 
as a function of the effective pump $\eta$ yields the threshold $\eta_c$ of the self-ordering.

Without loss of generality, consider $\alpha_0=0$ and the real condensate wavefunctions $\psi_{0\tau}=\sqrt{Nf_\tau}$,
where $f_\tau=N_\tau/N$ is the fraction of the atoms in state $\tau$ and the sum of the two condensate
wavefunctions is normalized to the total number of atoms, $\sum_\tau\int_{0}^{\lambda_c} \psi_{0\tau}^2dx/\lambda_c=N$.
The matrix $\mathbf{M}_{\rm B}$ can be readily obtained 
\begin{align} \label{eq:M-matrix}
\mathbf{M}_{\rm B}=
\begin{pmatrix}
\delta_c-i\kappa & 0 & \sqrt{N}\eta\sqrt{f_\uparrow}\mathcal{I} & \sqrt{N}\eta\sqrt{f_\uparrow}\mathcal{I} & 
\sqrt{N}\eta\sqrt{f_\downarrow}\mathcal{I} & \sqrt{N}\eta\sqrt{f_\downarrow}\mathcal{I} \\
0 & -(\delta_c+i\kappa) & -\sqrt{N}\eta\sqrt{f_\uparrow}\mathcal{I} & -\sqrt{N}\eta\sqrt{f_\uparrow}\mathcal{I} & 
-\sqrt{N}\eta\sqrt{f_\downarrow}\mathcal{I} & -\sqrt{N}\eta\sqrt{f_\downarrow}\mathcal{I} \\
\sqrt{N}\eta\sqrt{f_\uparrow} \cos(k_cx) & \sqrt{N}\eta\sqrt{f_\uparrow} \cos(k_cx) &
-\frac{\hbar}{2m}\frac{\partial^2}{\partial x^2} & 0 & 0 & 0 \\
-\sqrt{N}\eta\sqrt{f_\uparrow} \cos(k_cx) & -\sqrt{N}\eta\sqrt{f_\uparrow} \cos(k_cx) &
0 & \frac{\hbar}{2m}\frac{\partial^2}{\partial x^2} & 0 & 0 \\
\sqrt{N}\eta\sqrt{f_\downarrow} \cos(k_cx) & \sqrt{N}\eta\sqrt{f_\downarrow} \cos(k_cx) &
0 & 0 & -\frac{\hbar}{2m}\frac{\partial^2}{\partial x^2} & 0 \\
-\sqrt{N}\eta\sqrt{f_\downarrow} \cos(k_cx) & -\sqrt{N}\eta\sqrt{f_\downarrow} \cos(k_cx) &
0 & 0 & 0 & \frac{\hbar}{2m}\frac{\partial^2}{\partial x^2} 
\end{pmatrix},
\end{align}
where we have substituted the chemical potentials $\mu_\downarrow=0$ 
and $\mu_\uparrow=\tilde{\delta}$, and introduced
$\delta_c\equiv-\Delta_c+N\sum_\tau f_\tau U_\tau/2$ and the integral
operator $\mathcal{I}$,
\begin{align}
\mathcal{I}\xi=\frac{1}{\lambda_c}\int_{0}^{\lambda_c} \cos(k_cx)\xi dx.
\end{align}
Due to the form of the integral operator $\mathcal{I}$, only condensate fluctuations 
$\delta\psi_{\pm\tau}\propto\cos(k_cx)$ couple to the cavity-field fluctuation.
Therefore, we recast the matrix $\mathbf{M}_{\rm B}$ in the restricted subspace,
\begin{align} \label{eq:M-matrix-restricted}
\tilde{\mathbf{M}}_{\rm B}=
\begin{pmatrix}
\delta_c-i\kappa & 0 & \sqrt{N}\eta\sqrt{f_\uparrow}/2 & \sqrt{N}\eta\sqrt{f_\uparrow}/2& 
\sqrt{N}\eta\sqrt{f_\downarrow}/2 & \sqrt{N}\eta\sqrt{f_\downarrow}/2\\
0 & -(\delta_c+i\kappa) & -\sqrt{N}\eta\sqrt{f_\uparrow}/2 & -\sqrt{N}\eta\sqrt{f_\uparrow}/2& 
-\sqrt{N}\eta\sqrt{f_\downarrow}/2 & -\sqrt{N}\eta\sqrt{f_\downarrow}/2 \\
\sqrt{N}\eta\sqrt{f_\uparrow}  & \sqrt{N}\eta\sqrt{f_\uparrow}  &
\omega_r & 0 & 0 & 0 \\
-\sqrt{N}\eta\sqrt{f_\uparrow} & -\sqrt{N}\eta\sqrt{f_\uparrow}  &
0 & -\omega_r & 0 & 0 \\
\sqrt{N}\eta\sqrt{f_\downarrow} & \sqrt{N}\eta\sqrt{f_\downarrow}  &
0 & 0 & \omega_r & 0 \\
-\sqrt{N}\eta\sqrt{f_\downarrow} & -\sqrt{N}\eta\sqrt{f_\downarrow}  &
0 & 0 & 0 & -\omega_r
\end{pmatrix}.
\end{align}
The eigenvalues $\omega$ of $\tilde{\mathbf{M}}_{\rm B}$ is obtained via the 
sixth order characteristic equation $\text{Det}(\tilde{\mathbf{M}}_{\rm B}-\omega I_{6\times 6})=0$:
\begin{align}
(\omega^2-\omega_r^2)\left\{(\omega^2-\omega_r^2)[\delta_c^2+(i\omega-\kappa)^2]
+2\delta_c\omega_rN\eta^2\right\}=0.
\end{align}
The zero frequency solution $\omega=0$ yields the self-ordering threshold, 
\begin{align}
\sqrt{N}\eta_c=\sqrt{\frac{\delta_c^2+\kappa^2}{2\delta_c}}\sqrt{\omega_r}.
\end{align}


\end{document}